\documentclass[twocolumn,journal]{IEEEtran}
\usepackage[T1]{fontenc}
\usepackage[latin9]{inputenc}
\usepackage{float}
\usepackage{textcomp}
\usepackage{amsmath}
\usepackage{graphicx}
\usepackage[unicode=true,
 bookmarks=true,bookmarksnumbered=true,bookmarksopen=true,bookmarksopenlevel=1,
 breaklinks=false,pdfborder={0 0 0},pdfborderstyle={},backref=false,colorlinks=false]
 {hyperref}
\hypersetup{pdftitle={Your Title},
 pdfauthor={Your Name},
 pdfpagelayout=OneColumn, pdfnewwindow=true, pdfstartview=XYZ, plainpages=false}

\makeatletter

\usepackage[caption=false,font=footnotesize]{subfig}

\makeatother

\begin{document}
\title{To Study the Effect of Boundary Conditions and Disorder in Spin Chain Systems
Using Quantum Computers}
\author{M. Arsalan Ali\thanks{M. Arsalan Ali, Quaid-I-Azam Islamabad, Pakistan. e-mail: arsalanali2162@gmail.com}}

\maketitle
Condensed matter physics plays a crucial role in modern scientific
research and technological advancements, providing insights into the
the behavior of materials and their fundamental properties. Understanding
complex phenomena and systems in condensed matter physics pose significant
challenges due to their inherent intricacies. Over the years, computational
approaches have been pivotal in unraveling the mysteries of condensed
matter physics, but they face limitations when dealing with large-scale
systems and simulating quantum effects accurately. Quantum simulation
and quantum computation techniques have emerged as promising tools
for addressing these limitations, offering the potential to revolutionize
our understanding of condensed matter physics. In this paper, we focus
on the simulation of Anderson localization in the Heisenberg spin chain
systems and explore the effects of disorder on closed and open chain
systems using quantum computers.

\IEEEpeerreviewmaketitle{}

\section{Introduction}

Condensed matter physics investigates the properties and behaviors
of materials in various phases, ranging from solids and liquids to
complex electronic systems\cite{key-1}. The understanding of condensed
matter physics has played a pivotal role in enabling significant advancements
in nanotechnology, superconductivity\cite{key-2}, and quantum information
processing\cite{key-3}. Despite the remarkable progress made in the
field, understanding complex phenomena in condensed matter physics
remains a challenging task. These systems often exhibit emergent behavior,
where collective interactions between a large number of particles
give rise to novel properties. The interplay between different energy
scales, the presence of quantum fluctuations, and the effect of disorder
complicate the theoretical description of these systems\cite{key-4}
. Conventional computational techniques face limitations in efficiently
simulating such complex systems, hindering our ability to comprehend
their intricate behavior.

Various computational techniques\cite{key-5,key-6} have been employed
to study complex problems in condensed matter physics. Classical numerical
methods, such as Monte Carlo simulations\cite{key-7} and density
functional theory\cite{key-24}, have been widely used. While these
techniques have provided valuable insights, they are often limited
in their ability to handle large-scale systems or accurately capture
quantum effects. This necessitates exploring innovative computational
approaches, such as quantum simulation and quantum computation \cite{key-28}.

Quantum simulation\cite{key-29} involves emulating the behavior of
quantum systems using other controllable quantum systems, enabling
the study of complex quantum phenomena that are challenging to analyze
classically. Quantum computation, on the other hand, leverages quantum
mechanical principles to perform computations more efficiently than
classical computers in certain cases. By harnessing the principles
of superposition, entanglement, quantum interference, quantum
systems can potentially simulate and compute complex quantum phenomena
with unprecedented accuracy and efficiency\cite{key-30}.Quantum systems
possess unique properties that make them suitable for simulating complex
phenomena in condensed matter physics. These systems can exhibit long-range
entanglement, allowing for the exploration of quantum correlations
that play a vital role in understanding materials' properties. Moreover,
quantum simulators can be engineered to mimic specific Hamiltonian
and explore quantum phase transitions, quantum magnetism, and quantum
transport phenomena\cite{key-31}. The realm of quantum computers
opens up exciting possibilities, as their capacity to execute quantum
algorithms grant them the potential to efficiently tackle complex
problems.

Several companies\cite{key-12,key-13,key-14} are actively involved
in the development of quantum computing technologies. IBM Quantum
is one of the industry leaders, offering researchers and developers
access to quantum processors and software tools through the IBM Quantum
Experience platform.IBM Quantum has made significant progress in recent
years, advancing the capabilities and accessibility of quantum computing.
The platform provides cloud-based access to a range of quantum processors,
allowing researchers to design and execute quantum simulations. IBM
Quantum has continually improved the coherence and connectivity of
their quantum processors, enabling more complex simulations and computations.
Moreover, the platform offers a suite of programming tools and software
development kits (SDKs) to facilitate the implementation of quantum
algorithms.

This research paper aims to investigate the phenomenon of Anderson
localization in Heisenberg spin chain systems using quantum simulation
and computation techniques. Anderson localization\cite{key-15,key-16}
refers to the absence of electronic transport in disordered systems,
which has significant implications for condensed matter physics. The
the study focuses on the effect of defects on both closed and open chain
systems, shedding light on the behavior of quantum particles in disordered
environments. Quantum simulation methods provide a powerful tool for
studying such complex phenomena, and this paper presents a comprehensive
analysis of the results obtained through simulations on IBM's quantum
processors.

\section{Results}

Consider N spins 1/2 system, initially prepared in either $|\downarrow\downarrow\downarrow\uparrow\uparrow\uparrow>$
or Néel state$|\uparrow\downarrow\uparrow\downarrow\uparrow>$.\textbf{
}The evolution of time will be dictated by a Hamiltonian characterized
by the following form:

\textbf{
\begin{equation}
H_{1/2}=-g_{xy}\sum_{k=1}^{m-1}\left(\sigma_{k}^{x}\sigma_{k+1}^{x}+\sigma_{k}^{y}\sigma_{k+1}^{y}\right)+\sum_{k=1}^{m}\left(h_{k}\sigma_{k}^{z}\right)\label{eqn 01}
\end{equation}
}

\noindent with $g_{xy}>0$ represents nearest neighbor interaction
between $x(\sigma^{x})$ and $y(\sigma^{y})$ spins whereas $h_{k}$
represents an external field that couples to Z-spins at site $k$
and $\sigma_{k}^{\alpha}$ are the Pauli matrices with eigenvalues
\textpm 1. 

\noindent On basis of parameter $h_{k}\neq0$, we get following spin
chain system. Disordered $XX$ chain for $g_{xy}>0$ and $h_{k}$
where $h_{k}=[-h,h]$. The observable of interest is staggered
magnetization, which is defined as 

\begin{equation}
M_{s}(t)=\frac{1}{m}\sum_{k}(-1)^{k}<\sigma_{k}^{z}>\label{eqn 02}
\end{equation}

\noindent Here $m$ is the number of qubits. The choice of this observable
is necessitated by the realization that for a system with long range
antiferromagnetic order, the staggered magnetization is non-zero
and can be treated as the order parameter. Fortunately, $M_{s}(t)$
is in $z-basis$, so we only need post-process calculations on measured
qubits to compute the staggered magnetization. The expectation value
of spin operator Pauli-Z has value 1 for state $|0>$ and value 0
for state $|1>$. After measuring the qubit at the quantum back end, we
map the qubit with state $|0>$ to a value 1 and qubit with state
$|1>$ to a value of 1 to find the required observable which is staggered
magnetization in our case.

To simulate many body systems, IBM makes available different types
of simulators. Some of them are noise-free while others have noise
errors. In this regard, Aer is an ideal simulator with zero noise
while qasm simulator has some noise errors. We will compare the results
obtained on the quantum computer with those from qasm simulator. 

\noindent First of all Consider the disorder $XX$ chain, $H_{1/2}=-g_{xy}\sum_{k=1}^{m-1}\left(\sigma_{k}^{x}\sigma_{k+1}^{x}+\sigma_{k}^{y}\sigma_{k+1}^{y}\right)+\sum_{k=1}^{m}\left(h_{k}\sigma_{k}^{z}\right)$
with $g_{xy}>0$ and $h_{k}=0$ and a Neel state $\psi(t=0)=|\downarrow\uparrow\downarrow\uparrow>$
as an initial state to compute the staggered magnetization to predict
the behavior of the system.

\noindent As the expectation value of spin operator Pauli-Z has value
1 for state $|\uparrow>$ 

\[
\sigma_{z}|\uparrow>=1|\uparrow>
\]

\noindent and value -1 for state $|\downarrow>$

\[
\sigma_{z}|\downarrow>=-1|\downarrow>.
\]

\noindent From Eqn.(\ref{eqn 02}) and for Neel state $\psi(t=0)=|\downarrow\uparrow\downarrow\uparrow>$,\textbf{
}the staggered magnetization $M_{s}(t=0)$

\textbf{
\begin{equation}
M_{s}(t=0)=\frac{1}{m}\sum_{k}(-1)^{k}<\sigma_{k}^{z}>\label{eqn 03}
\end{equation}
}

{\footnotesize{}
\[
M_{s}(t=0)=\frac{1}{4}[(-1)^{1}<\downarrow\uparrow\downarrow\uparrow|\sigma_{z}^{1}|\downarrow\uparrow\downarrow\uparrow>+(-1)^{2}<\downarrow\uparrow\downarrow\uparrow|\sigma_{z}^{2}|\downarrow\uparrow\downarrow\uparrow>
\]
}{\footnotesize\par}

\[
+(-1)^{3}<\downarrow\uparrow\downarrow\uparrow|\sigma_{z}^{3}|\downarrow\uparrow\downarrow\uparrow>+(-1)^{4}<\downarrow\uparrow\downarrow\uparrow|\sigma_{z}^{4}|\downarrow\uparrow\downarrow\uparrow>]
\]

{\small{}
\[
M_{s}(t=0)=\frac{1}{4}\left[(-1)^{1}(-1)+(-1)^{2}(1)+(-1)^{3}(-1)+(-1)^{4}(1)\right]
\]
}{\small\par}

\[
M_{s}(t=0)=\frac{1}{4}\left[1+1+1+1\right]
\]

\[
M_{s}(t=0)=1
\]
Hence we expect $M_{s}(t=0)$=1 , where the initial state is Neel
state. 

\noindent For the disorder $XX$ chain the Hamiltonian is given as
\[
H_{1/2}=-g_{xy}\sum_{k=1}^{m-1}\left(\sigma_{k}^{x}\sigma_{k+1}^{x}+\sigma_{k}^{y}\sigma_{k+1}^{y}\right)+\sum_{k=1}^{m}\left(h_{k}\sigma_{k}^{z}\right)
\]

\noindent At later times ($t>0$), the initial state will evolve under
the $XX$ Hamiltonian given above. The staggered magnetization will
no longer have its initial value of 1 because the orientation of spins
will no longer be purely along the $z-axis$ but will begin to orient
in the $x-y$ plane. Hence staggered magnetization will approach zero
at later times.

\subsection{Effect of Disorder $h_{k}$ and Boundary Condition}

\subsubsection{Open Chain}

The Hamiltonian in given as 

\[
H_{1/2}=-g_{xy}\sum_{k=1}^{m-1}\left(\sigma_{k}^{x}\sigma_{k+1}^{x}+\sigma_{k}^{y}\sigma_{k+1}^{y}\right)+\sum_{k=1}^{m}\left(h_{k}\sigma_{k}^{z}\right)
\]

\begin{figure}[H]
\begin{minipage}[t]{0.45\columnwidth}%
\includegraphics[scale=0.4]{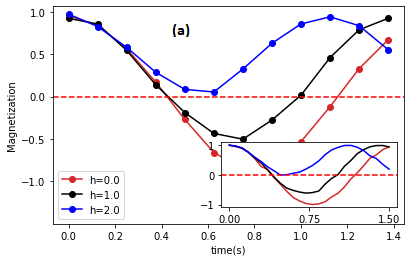}%
\end{minipage}\hfill{}%
\begin{minipage}[t]{0.45\columnwidth}%
\includegraphics[scale=0.4]{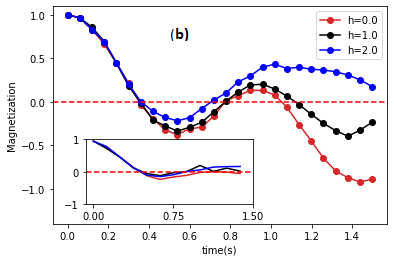}%
\end{minipage}\medskip{}
\noindent\begin{minipage}[c][1\totalheight][t]{1.1\columnwidth}%
\begin{flushleft}
\hspace{1cm}\includegraphics[scale=0.45]{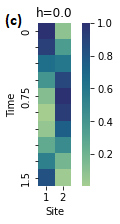}\includegraphics[scale=0.45]{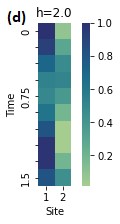}\includegraphics[scale=0.45]{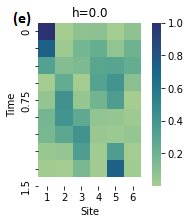}\includegraphics[scale=0.45]{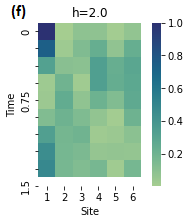}
\par\end{flushleft}%
\end{minipage}\caption[Disordered $XX$ open chain system for $g_{xy}=1$ and $h_{k}$.]{\label{fig 01}Disordered $XX$ Open chain system when $g_{xy}=1$
and $h_{k}$. Fig.(a) Magnetization for 2 sites. Fig.(b) Magnetization
for 4-sites. Fig.(c-e) Time and Disorder-Dependent Probability for$m=2$
and $m=4$Sites}
\end{figure}
\begin{figure}[H]
\begin{minipage}[t]{0.45\columnwidth}%
\includegraphics[scale=0.4]{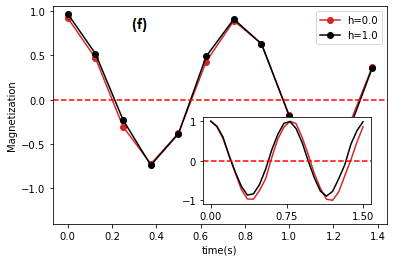}%
\end{minipage}\hspace{0.4cm}%
\begin{minipage}[t]{0.45\columnwidth}%
\includegraphics[scale=0.4]{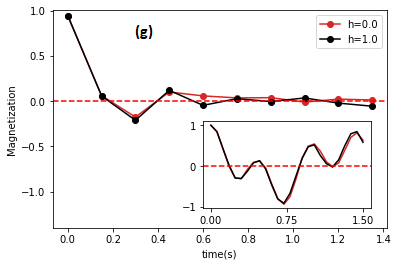}%
\end{minipage}\medskip{}
\noindent\begin{minipage}[t]{1.1\columnwidth}%
\hspace{1cm}\includegraphics[scale=0.45]{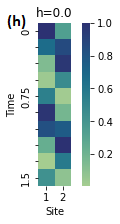}\includegraphics[scale=0.45]{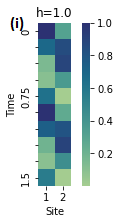}\includegraphics[scale=0.45]{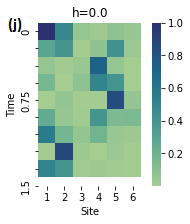}\includegraphics[scale=0.45]{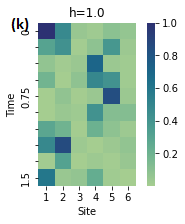}%
\end{minipage}
\textbf{\caption[Disordered $XX$ close chain system for $g_{xy}=1$ and $h_{k}$.]{\label{fig 02}Disordered $XX$ closed chain system when $g_{xy}=1$
and $h_{k}$. Fig.(f) Magnetization for 2 sites. Fig.(g) Magnetization
for 4 sites. Fig.(h-k) Time and Disorder-Dependent Probability in
a Closed Chain System for $m=2$ and $m=4$Sites}
}
\end{figure}

\noindent In Fig.(\ref{fig 01}a and b) we plotted staggered magnetization
for different values of $h_{k}$: the black line represent the staggered
magnetization for $h_{k}=0.0$, the red line for $h_{k}=0.5$and blue
line for $h_{k}=1$. In Fig.(\ref{fig 01}) the results in a small window
were simulated on IBM quantum devices while the solid line with
filled circles represents qasm simulator results for $XX$ chain. In
the Fig. (\ref{fig 01}(a)) the $XX$ chain model is simulated for
2-sites and in Fig.(\ref{fig 01}(b)) for 4-sites. In Fig.(\ref{fig 01})
we can see that the results of the IBM quantum computer deviate from the
qasm simulator as $\delta t>1$ because the available quantum devices
have a high error rate and particularly decoherence of qubits is high
for longer times. The deviation of the results becomes more noticeable
for the 2-site system, as the number of qubits rises, leading to increased
decoherence. In all three cases, the magnetization's time evolution
exhibits oscillatory behavior for both the 2-spin and 4-spin systems.\textbf{
}The explanation of oscillations in $M_{s}(t)$ has been given above
which holds in the presence of disorder. It is well established that
disorder leads to Anderson localization in 1D systems such as spin
chains under consideration and halts the dynamics. It is expected
that relaxation of staggered magnetization $M_{s}(t)$ will be slowed,
as seen in our plots and has been observed in recent past research\cite {key-17}.
Upon measuring a quantum system, its state undergoes a collapse, resulting
in a single basis vector. The probabilities of obtaining a basis state
upon measuring $|\psi(t)>$ are determined by the absolute squares
of the coefficients. By analyzing the probability plots, we can validate
our predictions regarding the impact of the disorder on the system's
evolution. In Fig.(\ref{fig 01}(c-f)), we can see that for small
values of disorder $h$ and open chain system nearly all states are
involved in the dynamics. Additionally, the intermediate states also
contribute to the overall behavior of the system. As the value of $h$
increases, the system requires more time to evolve from the initial
state |$\downarrow\uparrow$\textrangle{} to the final state $|\uparrow\downarrow\rangle$
and similarly, for four spin system, the system requires more time to
evolve from the initial state $|\downarrow\uparrow\downarrow\uparrow>$
to the final state $|\uparrow\downarrow\uparrow\downarrow>$. While
for the high value of disorder $h$, the probability of the intermediate
states will decrease and approaches zero. This implies that these
intermediate states are merely virtual states and not physically realized
in this particular case. They play a theoretical role in the dynamics
but do not manifest as observable outcomes.

\subsubsection{Closed Chain}

If we have a closed chain rather than an open chain, then the system
takes less time to evolve from the first state to the last state no
matter how large disorder $h_{k}$ is. This is so because now there
are two choices to get from the first state to last in case of closed
chain. It is well established that disorder leads to Anderson's localization
in 1D systems but for closed chain the system is not localized as
shown in the Fig.(\ref{fig 02}).

\section{Quantum Circuit}

\subsection{Quantum Bit}

The qubit state is given as
\begin{center}
$|0>=\left(\begin{array}{c}
1\\
0
\end{array}\right)$\hspace{3cm} $|1>=\left(\begin{array}{c}
0\\
1
\end{array}\right)$
\par\end{center}

\noindent We can visualization qubit states $|0>$and $|1>$ with
the help of Bloch sphere:

\begin{figure}[H]
\begin{minipage}[t]{0.35\columnwidth}%
\begin{flushright}
\includegraphics[scale=0.15]{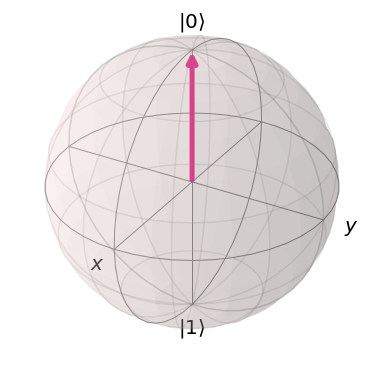}
\par\end{flushright}%
\end{minipage}\hspace{3cm}%
\begin{minipage}[t]{0.35\columnwidth}%
\begin{flushleft}
\includegraphics[scale=0.15]{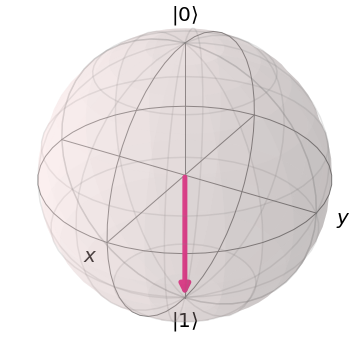}
\par\end{flushleft}%
\end{minipage}\caption{Qubit States$|0>$ and $|1>$ on Bloch Sphere}
\end{figure}

\subsection{Superposition}

Quantum bits are different from classical bits in terms of allowed
states.Classical bits can have only two states $|0>$ and $|1>$ but
quantum mechanics allows the qubit to take any value which is a coherent
superposition of $|0>$ and $|1>$ states.Hence, a qubit can be expressed
as a superposition\cite{key-10}, a combination of both |0| and |1|
states, as elucidated in reference. For instance, 

\textbf{
\begin{equation}
|a>=\frac{1}{\sqrt{2}}(|0>+|1>)\label{eqn 04}
\end{equation}
}

\noindent In the above state the probability of state $|0>$ and $|1>$
are equal which is $\frac{1}{2}$. On a Bloch sphere the state $|a>$
is represented as

\noindent 
\begin{figure}[H]
\begin{centering}
\begin{minipage}[t]{0.75\columnwidth}%
\begin{center}
\includegraphics[scale=0.35]{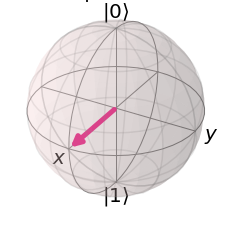}
\par\end{center}%
\end{minipage}
\par\end{centering}
\caption{Superposition of state $|0>$ and $|1>$ with equal probability on
Bloch Sphere}
\end{figure}

\noindent By using the superposition of $|0>$ and $|1>$ state we
can also define a number of different states which will be useful
when we discuss complex circuits:

{\footnotesize{}
\[
|b>=\frac{1}{\sqrt{2}}(|0>-|1>)\qquad|c>=\frac{1}{\sqrt{2}}(|0>+\iota|1>
\]
}{\footnotesize\par}

{\footnotesize{}
\[
|d>=\frac{3}{\sqrt{4}}(|0>+\sqrt{\frac{1}{4}}|1>)\qquad|e>=\frac{\sqrt{3}}{2}(|0>+\frac{\iota-1}{\sqrt{8}}|1>)
\]
}{\footnotesize\par}

\begin{figure}[H]
\begin{minipage}[t]{0.22\columnwidth}%
\begin{center}
\includegraphics[scale=0.35]{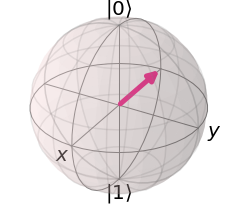}
\par\end{center}%
\end{minipage}\hfill{}%
\begin{minipage}[t]{0.22\columnwidth}%
\begin{center}
\includegraphics[scale=0.35]{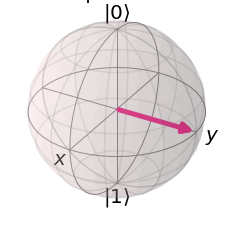}
\par\end{center}%
\end{minipage}\hfill{}%
\begin{minipage}[t]{0.22\columnwidth}%
\begin{center}
\includegraphics[scale=0.35]{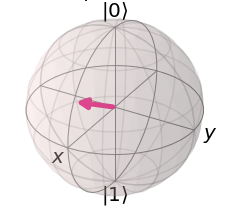}
\par\end{center}%
\end{minipage}\hfill{}%
\begin{minipage}[t]{0.22\columnwidth}%
\begin{center}
\includegraphics[scale=0.35]{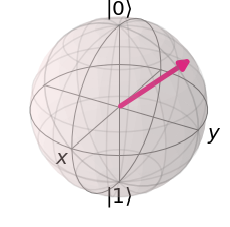}
\par\end{center}%
\end{minipage}
\centering{}\caption{Superposition states $|b>$,$|c>$, $|d>$ and $|e>$ on Bloch sphere}
\end{figure}

The details of how we can measure single and multiple qubit states are
given in Appendix I.

\subsection{Gates}

First, we discuss Pauli gates
\begin{center}
\textbf{$\sigma^{x}=\left(\begin{array}{cc}
0 & 1\\
1 & 0
\end{array}\right)$$\hspace{0.2cm}\sigma^{y}=\left(\begin{array}{cc}
0 & -\iota\\
\iota & 0
\end{array}\right)\hspace{0.2cm}$$\sigma^{z}=\left(\begin{array}{cc}
1 & 0\\
0 & -1
\end{array}\right)$}
\par\end{center}

\noindent Pauli x gate rotate the state by$180^{o}$, y gate is a
combination of bit-flip and phase flip and z-gate only flips the phase.

\textbf{}
\begin{figure}[H]
\textbf{}%
\begin{minipage}[t]{0.33\columnwidth}%
\begin{center}
\textbf{\includegraphics[scale=0.5]{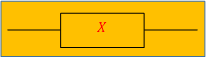}}
\par\end{center}%
\end{minipage}\textbf{\hfill{}}%
\begin{minipage}[t]{0.33\columnwidth}%
\begin{center}
\textbf{\includegraphics[scale=0.5]{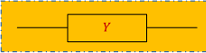}}
\par\end{center}%
\end{minipage}\textbf{\hfill{}}%
\begin{minipage}[t]{0.33\columnwidth}%
\begin{center}
\textbf{\includegraphics[scale=0.5]{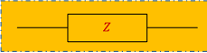}}
\par\end{center}%
\end{minipage}\textbf{\caption{Pauli X, Y and Z gates}
}
\end{figure}

\noindent Using rotation gates, we have the ability to rotate a state
by an angle $\phi$ around the$x,y,$or$z-axis,$ thereby obtaining
the rotated state. The following rotation gates implement rotations
about the $x,y,$ and $z-axis$:
\begin{center}
\textbf{$R^{x}(\phi)=e^{-i\frac{\phi\sigma^{x}}{2}}$\hspace{0.1cm}
$R^{y}(\phi)=e^{-i\frac{\phi\sigma^{y}}{2}}$\hspace{0.1cm}$R^{z}(\phi)=e^{-i\frac{\phi\sigma^{z}}{2}}$}
\par\end{center}

\textbf{}
\begin{figure}[H]
\textbf{}%
\begin{minipage}[t]{0.33\columnwidth}%
\begin{center}
\textbf{\includegraphics[scale=0.5]{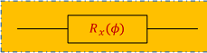}}
\par\end{center}%
\end{minipage}\textbf{\hfill{}}%
\begin{minipage}[t]{0.33\columnwidth}%
\begin{center}
\textbf{\includegraphics[scale=0.5]{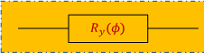}}
\par\end{center}%
\end{minipage}\textbf{\hfill{}}%
\begin{minipage}[t]{0.33\columnwidth}%
\begin{center}
\textbf{\includegraphics[scale=0.5]{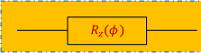}}
\par\end{center}%
\end{minipage}\caption{Rotation $R^{x}(\phi),R^{y}(\phi)$ and $R^{z}(\phi)$ gates}
\end{figure}

\noindent The Hadamard gate is a single-qubit gate, defined as follows
as:\textbf{}
\textbf{
\[
H=\frac{1}{\sqrt{2}}\left(\begin{array}{cc}
1 & 1\\
1 & -1
\end{array}\right)
\]
}

\noindent Hadamard gate acting on either $|0>$ or $|1>$ state creates
a state which is an equal superposition of $|0>$ and $|1>$. This
will turn out to be very useful in quantum computation.
\begin{center}
$H|0>=\frac{1}{\sqrt{2}}(|0>+|1>)\hspace{0.5cm}$$H|1>=\frac{1}{\sqrt{2}}(|0>-|1>)$
\par\end{center}

\noindent Another improtant single qubit gate is phase gate. It will
only change the phase of the state either apply on $|0>$ or $|1>$

\[
S=\left(\begin{array}{cc}
1 & 0\\
0 & e^{i\lambda}
\end{array}\right)
\]

\begin{figure}[tbh]
\begin{minipage}[t]{0.5\columnwidth}%
\begin{center}
\includegraphics[scale=0.5]{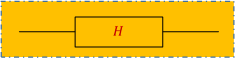}
\par\end{center}%
\end{minipage}%
\begin{minipage}[t]{0.49\columnwidth}%
\begin{center}
\includegraphics[scale=0.5]{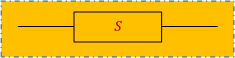}
\par\end{center}%
\end{minipage}

\caption{Circuit for Hadamard and Phase gate}
\end{figure}

\noindent The first multi qubit gate is $CNOT$ gate but we can reverse
the $CNOT$ gate by applying circuit as shown in fig(\ref{fig 03})
\begin{center}
{\footnotesize{}$CNOT=\left(\begin{array}{cccc}
1 & 0 & 0 & 0\\
0 & 1 & 0 & 0\\
0 & 0 & 0 & 1\\
0 & 0 & 1 & 0
\end{array}\right)$}{\footnotesize\par}
\par\end{center}

\begin{center}
{\footnotesize{}$RCNOT=\left(\begin{array}{cccc}
1 & 0 & 0 & 0\\
0 & 0 & 0 & 1\\
0 & 0 & 1 & 0\\
0 & 1 & 0 & 0
\end{array}\right)$}
\begin{figure}[H]
\begin{minipage}[t]{0.4\columnwidth}%
\begin{flushright}
\includegraphics[scale=0.48]{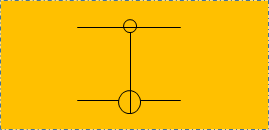}
\par\end{flushright}%
\end{minipage}\hspace{1.5cm}%
\begin{minipage}[t]{0.4\columnwidth}%
\includegraphics[scale=0.5]{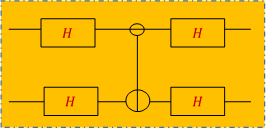}%
\end{minipage}

\caption{\label{fig 03}Circuit for $Cnot$ and $RCnot$ gates}
\end{figure}
\par\end{center}

\section{Mathematical model for Simulating Spin chain Hamiltonian}

The Hamiltonian for spin chains\cite{key-21} system is given as

\[
H_{1/2}=-g_{xy}\sum_{k=1}^{m-1}\left(\sigma_{k}^{x}\sigma_{k+1}^{x}+\sigma_{k}^{y}\sigma_{k+1}^{y}\right)+\sum_{k=1}^{m}\left(h_{k}\sigma_{k}^{z}\right)
\]

\noindent As our focus lies in dynamics, specifically the time evolution
of states, we construct the time evolution operator\textbf{ }

\begin{figure}[H]
\noindent\begin{minipage}[t]{1\columnwidth}%
\begin{center}
\includegraphics[scale=0.4]{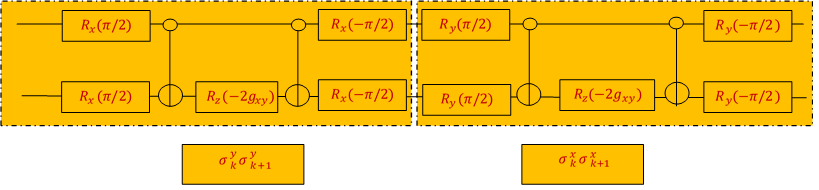}
\par\end{center}%
\end{minipage}\caption{\label{fig 04}Quantum circuit for the digital quantum simulation
of 2-qubit spin chain system}
\end{figure}

{\footnotesize{}
\begin{equation}
e^{-iH_{1/2}t}=\left(e^{-i\left[\sum_{k=1}^{m-1}\left\{ -g_{xy}\left(\sigma_{k}^{x}\sigma_{k+1}^{x}+\sigma_{k}^{y}\sigma_{k+1}^{y}\right)+h_{k}\left(h_{k}\sigma_{k}^{z}\right)\right\} \right]t}\right)\label{eqn 08}
\end{equation}
}{\footnotesize\par}

\noindent we can implement the required operation\cite{key-22} needed
in time evolution operator for the Hamiltonian for spin chain system
given in Eqn.(\ref{eqn 08}) as shown in Fig.(\ref{fig 04}). 

It is well-established that any n-qubit quantum computation can be
accomplished using a sequence of one-qubit and two-qubit quantum logic
gates. Nevertheless, finding the optimal circuit for a specific family
of gates, even for two-qubit gates, is a challenging task. This poses
an issue as quantum computation experimentalists can currently only
execute a limited number of gate operations within the coherence time
of their physical systems. Without an established procedure for optimal
quantum circuit design, experimentalists might encounter difficulties
in demonstrating certain quantum operations.\textbf{ }In Fig.(\ref{fig 04})
we can see that $6-CNot$ gates are required to solve two-qubit spin
chain system while in Fig. (\ref{fig 05}) which is given below, we
can see that $2-CNot$ gates are used to solve 2-qubit spin chain
system.\textbf{ }Indeed, in this context, we review a procedure for
constructing an optimal quantum circuit\cite{key-23} that enables
the realization of a general two-qubit quantum computation. The detail
of the procedure is given in Appendix II.

\begin{figure}[H]
\noindent\begin{minipage}[t]{1\columnwidth}%
\begin{center}
\includegraphics[scale=0.5]{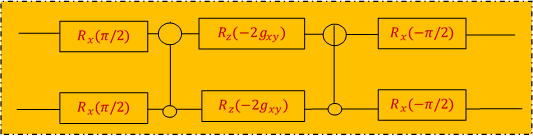}
\par\end{center}%
\end{minipage}\caption{\label{fig 05}Quantum circuit to implement $e^{i(\sigma_{k}^{x}\sigma_{k+1}^{x}+\sigma_{k}^{y}\sigma_{k+1}^{y})g_{xy}}$}
\end{figure}

\section{Trotter Decomposition}

To implement time evolution of states governed by $H_{1/2}$numerically,
we employ Trotter decomposition.

\noindent Using Trotter decomposition\cite{key-19} 

\textbf{\footnotesize{}
\begin{equation}
e^{-iH_{1/2}t}\approx\left(\prod_{k=1}^{n}e^{-iH_{1/2}\delta t/m}\right)^{m}\label{eqn 09}
\end{equation}
}{\footnotesize\par}

\noindent where $n$ is a number of terms. As we increase the number
of Trotter steps, the results of the Trotter approximation get closer
to the desired outcome. However, this improvement comes at the expense
of increasing the number of gates, leading to a longer circuit length\cite{key-20}. 

Let's consider a Hamiltonian of the form $H_{1/2}=\hat{O}+\hat{Q}$,
where $[\hat{O},\hat{Q}]\neq0$. Since these operators do not generally
commute, we can approximate the time evolution as follows:

\[
e^{-iH_{1/2}t}\approx e^{-i\hat{O}t}e^{-i\hat{Q}t}+O(t)
\]

This approximation can be naturally extended to Hamiltonian that
consists of more than two terms. To utilize this fact for tottered
evolution, we can follow the two steps outlined in the main text.
Let's now delve into the details of these steps more thoroughly. For
the Hamiltonian

\[
H_{1/2}=-g_{xy}\sum_{k=1}^{m-1}\left(\sigma_{k}^{x}\sigma_{k+1}^{x}+\sigma_{k}^{y}\sigma_{k+1}^{y}\right)+\sum_{k=1}^{m}\left(h_{k}\sigma_{k}^{z}\right)
\]

\noindent and 

\textbf{
\[
e^{-iH_{1/2}t}=\left(e^{-i\left[\sum_{k=1}^{m-1}\left\{ -g_{xy}\left(\sigma_{k}^{x}\sigma_{k+1}^{x}+\sigma_{k}^{y}\sigma_{k+1}^{y}\right)+h_{k}\left(h_{k}\sigma_{k}^{z}\right)\right\} \right]t}\right)
\]
}

\noindent We first define the operators

\textbf{$\hat{O_{k}}=e^{-ih_{k}\sigma_{k}^{z}\delta t}$, $\hat{Q_{k}}=e^{-i(-g_{xy}(\sigma_{k}^{x}\sigma_{k+1}^{x}+\sigma_{k}^{y}\sigma_{k+1}^{y}))\delta t}$}

\noindent Using $\hat{O_{k}}$ and$\hat{Q_{k}}$

\begin{figure}[H]
\textbf{}%
\noindent\begin{minipage}[t]{1\columnwidth}%
\begin{center}
\textbf{\includegraphics[scale=0.3]{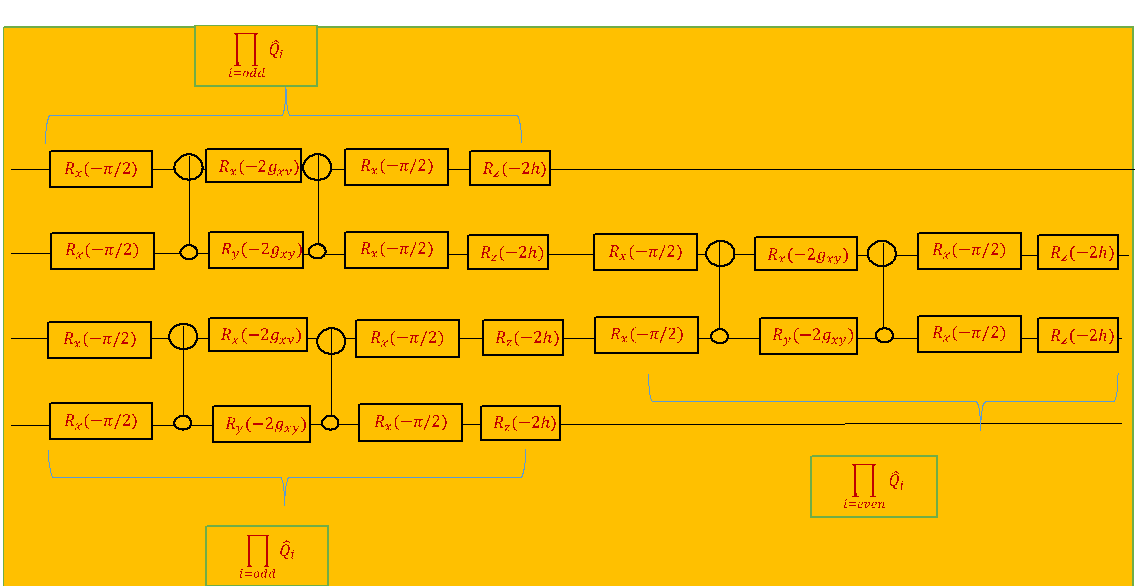}}
\par\end{center}%
\end{minipage}\textbf{\caption{\label{fig 06}Disordered $XX$ chain for $g_{xy}=1$ and $h_{k}$
where $h_{k}=[-h,h]$.}
}
\end{figure}

\textbf{\footnotesize{}
\begin{equation}
e^{-i\hat{Ht}}\approx\left(\prod_{k}\hat{O_{k}}\right)\left(\prod_{k=even}\hat{Q_{k}}\right)\left(\prod_{k=odd}\hat{Q_{k}}\right)+O(\delta t)\label{eqn 11}
\end{equation}
}{\footnotesize\par}

\noindent which shown in the Fig.(\ref{fig 06}).

\noindent For $4$ sites 

\textbf{\footnotesize{}
\[
\prod_{k}O_{k}=\left(e^{-ih_{1}\sigma_{1}^{z}\delta t}\right)\left(e^{-ih_{2}\sigma_{2}^{z}\delta t}\right)\left(e^{-ih_{3}\sigma_{3}^{z}\delta t}\right)\left(e^{-ih_{4}\sigma_{4}^{z}\delta t}\right)
\]
}{\footnotesize\par}

\noindent For $k$ is even

\textbf{
\[
\prod_{k=even}\hat{Q}=e^{ig_{xy}(\sigma_{2}^{x}\sigma_{3}^{x}+\sigma_{3}^{y}\sigma_{4}^{y}))\delta t}
\]
}

\noindent For $k$ is odd

\textbf{
\[
\prod_{k=odd}\hat{Q}=\left(e^{ig_{xy}(\sigma_{1}^{x}\sigma_{2}^{x}+\sigma_{1}^{y}\sigma_{2}^{y})\delta t}\right)\left(e^{ig_{xy}(\sigma_{3}^{x}\sigma_{4}^{x}+\sigma_{3}^{y}\sigma_{4}^{y})\delta t}\right)
\]
}

\noindent Quantum circuit to implement Eqn.(\ref{eqn 11}) operation
is shown in Fig.(\ref{fig 06})

\textbf{}

The detail of the algorithm to implement the circuit on IBM quantum experience
is given in

\section{Error Analysis}

In this paper, we have determined the time evolution of many-body
states of spin chains by employing quantum algorithms run on IBM quantum
computers. Currently, IBM quantum devices have noise errors due to multi-qubit gates and decoherence of qubits so on an hourly basis these machines
are calibrated; that is why the results and data obtained at certain
times may have high error rates as compared to other times/days. So
the real-time data of different quantum processors can be checked anytime\cite{key-18}.

\begin{figure}[H]
\begin{centering}
\includegraphics[scale=0.5]{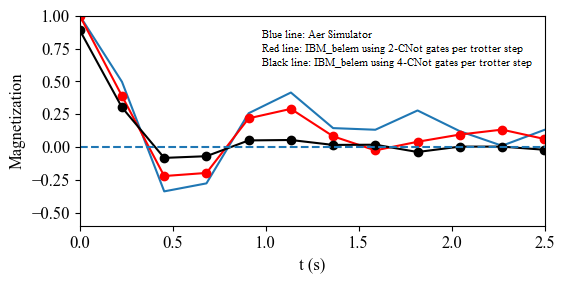}
\par\end{centering}
\caption[Scheme of quantum circuit for $4-CNot$ and $2-CNot$ gates]{\label{fig 07}Scheme of quantum circuit using: $4-CNot$ gates(black
line) per trotter step and $2-CNot$ gates (orange line) per trotter
step}
\end{figure}

\noindent For the time evolution of an initial state governed by the
Hamiltonian and to obtain an observable quantity we calculate the
observable which is staggered magnetization. We have discussed two
circuits: one is shown in fig(\ref{fig 04}) which has $4-CNot$ gates
and other circuit is shown in fig(\ref{fig 05}) which has $2-CNot$
gates for the disorder $XX$ chain. The results corresponding to both
quantum circuits are shown in Fig.(\ref{fig 07})

\noindent This analysis holds significant importance since one effective
method to reduce the error rate is by minimizing the number of gates\textbf{.}
Above Fig.(\ref{fig 07}) the black line represents quantum circuit
scheme which is using $4-CNot$ gates in one trotter step and the
the orange line represents $2-CNot$ gates in one trotter step while the
the blue line shows the results of an ideal simulator named Aer. To have
a better comparison we take $12-trotter$ steps for time evolution
of states otherwise due to the decoherence of qubits and an error rate of
$CNot$ gates the results deviated quickly from ideal results.

\noindent Black line: $4-CNot$ gates per trotter step and there are
total of 12 steps for time evolution which implies that the total
number of $CNot$ gates will be $48$.

\noindent Orange line: $2-CNot$ gates per trotter step and there
are a total of 12 steps which implies that the total number of $CNot$
gates will be $24$. 

\noindent We have employed the quantum circuit scheme which has a minimum
error due to fewer number of $CNot$ gates. This can be seen from
the above graphs since the orange line is closer to the ideal simulation
as compared to the black line. Now, we will study the magnetization for
three different cases of spin chains by using the quantum circuit
which has $(2-CNot)$ gates per trotter step.

\noindent \textbf{Experimental Realization: }Studying quantum spin
chain systems in a laboratory setting is a challenging exercise because
different parameters are present which are difficult to control. Ultra
cold atoms in optical lattices have proven to be an invaluable experimental
system for investigating spin chain systems. Their unique characteristics
allow for precise control and manipulation, offering a high degree
of contrast essential for initializing spin systems and studying their
dynamics. In this regard, the initialization of the many body spin state,
tunability of exchange interaction and time-resolved magnetization
have been demonstrated in ultra cold atoms systems(see \cite{key-25,key-26}
and reference therein).

\noindent \textbf{Future Outlook: }Work presented in this thesis can
be extended in several directions. One clear extension, even though
more challenging, is the extension of this work to 2D systems, both ferromagnetic
and anti-ferromagnetic\cite {key-28}. Another direction is driven
systems such as Floquet systems\cite{key-29,key-30}. As new and improved
quantum computers are developed more challenging problems can be attacked.
The work carried out in this thesis provides the perfect platform
for it.

\appendices{}

\section*{Appendix I}

\subsection*{Measurement }

\subsubsection*{Single Qubit states:}

In section 3.2, we have seen the following superposition state of
a qubit

\begin{equation}
|b>=\frac{1}{\sqrt{2}}(|0>-|1>).\label{eqn 05}
\end{equation}
\begin{figure}[H]
\begin{centering}
\includegraphics[scale=0.35]{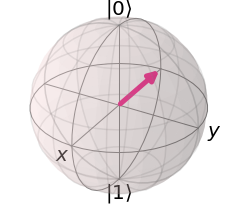}
\par\end{centering}
\caption{Qubit state $|b>=\frac{1}{\sqrt{2}}(|0>-|1>)$ lies half way between
north and south pole on Bloch sphere.}
\end{figure}

\noindent Quantum mechanics allows states to be superpositions but
when we measure a quantum state we get a defined state $|0>$ or $|1>$
with some probability. It means the superposition of qubits collapses to
a defined state when we measure it at the end of the computation. For example,
in the above qubit $|b>$ , if we measure it we get $|0>$state with
probability $1/2$ and $|1>$with probability $1/2$. Geometrically
the qubit lies halfway between the north and south pole of the Bloch sphere. 

\noindent As another example we calculate the probabilities of another
qubit:

{\footnotesize{}
\[
|e>=\frac{\sqrt{3}}{2}(|0>+\frac{\iota-1}{\sqrt{8}}|1>
\]
}{\footnotesize\par}

\noindent Probability \textbf{ }of state $|0>$ is given as

{\footnotesize{}
\[
|\frac{\sqrt{3}}{2}|^{2}=\frac{3}{4}
\]
}Probability of state $|1>$ is given as

{\footnotesize{}
\[
|\frac{\iota-1}{\sqrt{8}}|^{2}=\frac{2}{8}
\]
}{\footnotesize\par}

\noindent So on measurement $|0>$ is obtained with probability $\frac{3}{4}$
and the probability of $|1>$is $\frac{2}{8}$, which is shown in
Fig.(2.4)

\subsubsection*{Multiple Qubit States:}

In section 3.3 we have discussed single qubit states, the superposition
of states of single qubits, and how to measure these states. Now we
will move towards superposition and measurement of multiple qubits
states. In multiple qubits, we write the states of both qubits as a
tensor product, for example 

\[
|0>\otimes|0>)=|00>.
\]

\noindent For two qubit we get $2^{2}$ states 

\noindent 
\[
\{|00>,|01>,|10>,|11>\}
\]

\noindent The matrix representations of these states is

{\footnotesize{}
\[
|00>=|0>\otimes|0>)=\left(\begin{array}{c}
1\\
0
\end{array}\right)\otimes\left(\begin{array}{c}
1\\
0
\end{array}\right)=\left(\begin{array}{c}
1\\
0\\
0\\
0
\end{array}\right)
\]
}{\footnotesize\par}

{\footnotesize{}
\[
|01>=|0>\otimes|1>)=\left(\begin{array}{c}
1\\
0
\end{array}\right)\otimes\left(\begin{array}{c}
0\\
1
\end{array}\right)=\left(\begin{array}{c}
0\\
0\\
0\\
1
\end{array}\right)
\]
}{\footnotesize\par}

{\footnotesize{}
\[
|10>=|1>\otimes|0>)=\left(\begin{array}{c}
0\\
1
\end{array}\right)\otimes\left(\begin{array}{c}
1\\
0
\end{array}\right)=\left(\begin{array}{c}
0\\
0\\
1\\
0
\end{array}\right)
\]
}{\footnotesize\par}

{\footnotesize{}
\[
|11>=|1>\otimes|1>)=\left(\begin{array}{c}
0\\
1
\end{array}\right)\otimes\left(\begin{array}{c}
0\\
1
\end{array}\right)=\left(\begin{array}{c}
0\\
0\\
0\\
1
\end{array}\right).
\]
}{\footnotesize\par}

\noindent The superposition of these states is given as

\textbf{
\begin{equation}
|\phi_{2_{qubit-states}}>=\alpha_{0}|00>+\alpha_{1}|01>+\alpha_{2}|10>+\alpha_{3}|11>\label{eqn 06}
\end{equation}
}The probabilities of obtaining qubit states are as follows:

1. Probability of $|00|\text{is }|\alpha_{0}|^{2}$

2. Probability of $|01|\text{is }|\alpha_{1}|^{2}$

3. Probability of |$10|\text{is }|\alpha_{2}|^{2}$

4. Probability of $|11|\text{is}|\alpha_{3}|^{2}$

\noindent If we measure only the left qubit $|0>$ the probability
is given as $|\alpha_{0}|^{2}+|a_{1}|^{2}$.\textbf{ }Allow us to
consider an example utilizing the following two-qubit state:

\textbf{\footnotesize{}
\begin{equation}
|\gamma>=\frac{1}{\sqrt{7}}|00>+\frac{2}{\sqrt{7}}|01>+\frac{3}{\sqrt{7}}|10>+\frac{1}{\sqrt{7}}|11>\label{eqn 07}
\end{equation}
}{\footnotesize\par}

\noindent The probability of measuring $|0>$ is given as 

{\footnotesize{}
\[
|\frac{1}{\sqrt{7}}|^{2}+|\frac{2}{\sqrt{7}}|^{2}=\frac{3}{7}
\]
}{\footnotesize\par}

\noindent For three qubit we have $2^{3}$states

\noindent \textbf{\footnotesize{}
\[
\{|000>,|010>,|001>,|100>,|110>,|101>,|011>,|111>\}
\]
}{\footnotesize\par}

\noindent The superposition of these states is given as

\textbf{\small{}
\[
|\phi_{3_{qubit-states}}>=\beta_{0}|000>+\beta_{1}|010>+\beta_{2}|001>+\beta_{3}|100>
\]
}{\small\par}

\textbf{\small{}
\[
+\beta_{4}|110>+\beta_{5}|101>+\beta_{6}|011>+\beta_{7}|111>
\]
}{\small\par}

\noindent if we measure the qubit state $|000>$ its probability is
$|\beta_{0}|^{2}$, the probability of $|010>$ is $|\beta_{1}|^{2}$,
the probability of $|001>$ is $|\beta_{2}|^{2}$and so on.

\noindent For a system with $m$ qubits, there exist $2^{m}$ states.
The superposition state can be expressed as follows:

\textbf{
\[
|\phi_{m_{qubit-states}}>=\beta_{0}|000..m>+\beta_{1}|010....m>
\]
}

\textbf{
\[
+\beta_{m}|001...m>+..............=\left(\begin{array}{c}
\beta_{0}\\
\beta_{1}\\
....\\
....\\
....\\
\beta_{m}
\end{array}\right)
\]
}

\section*{Appendix II}

\noindent Consider

\noindent 
\[
N(\lambda,\gamma,\phi)=e^{i(\lambda\sigma_{k}^{x}\sigma_{k+1}^{x}+\gamma\sigma_{k}^{y}\sigma_{k+1}^{y}+\phi\sigma_{k}^{z}\sigma_{k+1}^{z})}
\]

For x and z terms only, $\gamma=0$

\noindent 
\[
N(\lambda,0,\phi)=e^{i(\lambda\sigma_{k}^{x}\sigma_{k+1}^{x}+\phi\sigma_{k}^{z}\sigma_{k+1}^{z})}
\]

\noindent As we have Magic matrix\cite{key-14}

{\footnotesize{}
\[
M=\frac{1}{\sqrt{2}}\left(\begin{array}{cccc}
1 & i & 0 & 0\\
0 & 0 & i & 1\\
0 & 0 & i & -1\\
1 & -i & 0 & 0
\end{array}\right)
\]
}
\begin{figure}[H]
\noindent\begin{minipage}[t]{1\columnwidth}%
\begin{center}
\includegraphics[scale=0.35]{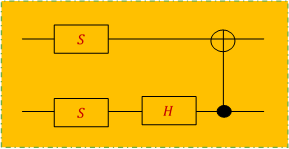}
\par\end{center}%
\end{minipage}\caption{\label{Fig10}Quantum circuit to implement magic basis}
\end{figure}

\noindent Using this matrix we can find that 

{\scriptsize{}
\begin{equation}
M^{\dagger}N(\lambda,0,\phi)M=e^{i\phi\sigma^{z}}\otimes e^{i\lambda\sigma^{z}}.\label{eqn 12}
\end{equation}
}{\scriptsize\par}

{\footnotesize{}$\implies$}{\footnotesize\par}

{\footnotesize{}
\begin{equation}
N(\lambda,0,\phi)=Me^{i\phi\sigma^{z}}\otimes e^{i\lambda\sigma^{z}}M^{\dagger}\label{eqn 13}
\end{equation}
}{\footnotesize\par}

\begin{figure}[H]
\noindent\begin{minipage}[t]{1\columnwidth}%
\begin{center}
\includegraphics[scale=0.4]{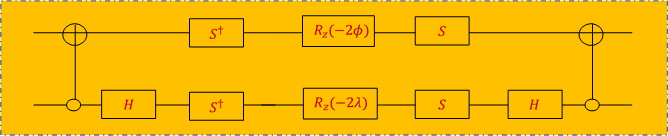}
\par\end{center}%
\end{minipage}\caption{\label{Fig11}Quantum circuit to implement $N(\lambda,0,\phi)=Me^{i\phi\sigma^{z}}\otimes e^{i\lambda\sigma^{z}}M^{\dagger}$}
\end{figure}

\noindent Quantum circuit to implement $Me^{i\phi\sigma^{z}}\otimes e^{i\lambda\sigma^{z}}M^{\dagger}$is
shown in Fig.(\ref{Fig11}) 

As $[S,R_{z}(\theta)]=0$ and $HR_{z}(\theta)H=R_{x}(\theta)$. Then
circuit reduced to

\begin{figure}[tbh]
\begin{centering}
\includegraphics[scale=0.4]{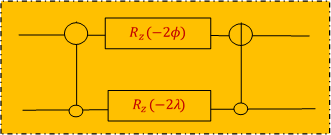}
\par\end{centering}
\caption{Quantum circuit to implement $N(\lambda,0,\phi)=Me^{i\phi\sigma^{z}}\otimes e^{i\lambda\sigma^{z}}M^{\dagger}$}
\end{figure}

\end{document}